# Isotropy-Optimized Contrastive Learning for Semantic Course Recommendation

Ali Khreis, Anthony Nasr, Yusuf Hilal
School of Electrical and Computer Engineering
University of Ottawa
Ottawa, Canada

*Abstract – **This paper presents a semantic course recommendation system for students using a self-supervised contrastive learning approach built upon BERT (Bidirectional Encoder Representations from Transformers). Traditional BERT embeddings suffer from anisotropic representation spaces, where course descriptions exhibit high cosine similarities regardless of semantic relevance. To address this limitation, we propose a contrastive learning framework with data augmentation and isotropy regularization that produces more discriminative embeddings. Our system processes student text queries and recommends Top-N relevant courses from a curated dataset of over 500 engineering courses across multiple faculties. Experimental results demonstrate that our fine-tuned model achieves improved embedding separation and more accurate course recommendations compared to vanilla BERT baselines.***

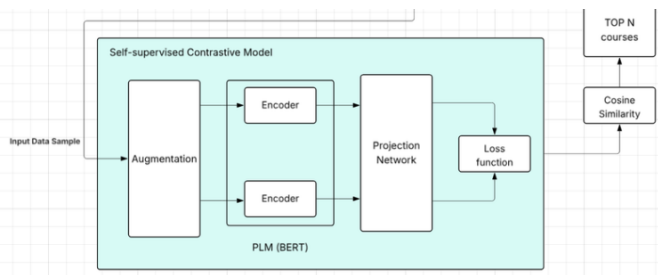

Fig. 1. Model Specific flowchart

## I. INTRODUCTION

Course selection is an important yet challenging task for engineering students, particularly at large institutions such as the University of Ottawa, where hundreds of courses are offered across multiple engineering disciplines. While program requirements and academic advising provide general guidance, students often struggle to identify courses that best match their individual interests and career goals. This challenge is amplified for new students or those pursuing interdisciplinary pathways, where little prior enrollment data is available.

Recommender systems have been widely used to personalize user experiences in domains such as e-commerce and online learning. Traditional course recommendation approaches typically rely on collaborative filtering, which infers preferences from historical enrollment data. Although effective in some settings, these methods suffer from cold-start and long-tail problems, making them poorly suited for academic environments where personalization based on content is essential.

Content-based recommendation methods address these limitations by leveraging course descriptions directly. Recent advances in natural language processing, particularly pretrained language models such as BERT, enable contextual semantic representations of text and allow both student interest statements and course descriptions to be embedded into a shared vector space. However, prior work has shown that vanilla BERT embeddings are highly anisotropic, causing unrelated texts to exhibit high cosine similarity and limiting their effectiveness for similarity-based retrieval tasks.

To overcome this limitation, we propose a **self-supervised contrastive learning framework** built on top of a pretrained BERT encoder for course recommendation. By using text-based data augmentation and contrastive objectives, the proposed method encourages semantically related inputs to cluster together while pushing unrelated samples apart. An additional isotropy regularization term is introduced to improve the geometric structure of the embedding space. This results in more discriminative representations that are well suited for cosine similarity–based ranking.

We apply the proposed approach to recommending engineering courses at the University of Ottawa based on free-form student interest statements. A curated dataset of engineering course descriptions and synthetic student queries is constructed, and the system is evaluated using standard retrieval metrics and embedding isotropy analysis. Experimental results show that the proposed model significantly outperforms vanilla BERT embeddings in both recommendation accuracy and embedding quality.

**The main contributions of this work are as follows:**

- A contrastive learning–based course recommendation system operating on natural language student input.
- A method to mitigate anisotropy in pretrained BERT embeddings using contrastive learning and isotropy regularization.
- An empirical evaluation demonstrating improved recommendation performance over vanilla BERT baselines.

## II. LITERATURE REVIEW

### A. Traditional Recommendation Approaches

Early course recommendation systems relied on collaborative filtering [1], which identifies patterns in student enrollment histories to suggest courses. While effective when sufficient data exists, these approaches suffer from cold-start problems for new students and fail to capture course content semantics.

Content-based methods using TF-IDF vectorization [2] address some limitations by analyzing course descriptions directly. However, these bag-of-words approaches cannot capture semantic relationships between terms or handle synonyms effectively.

### B. Deep Learning in Recommendation Systems

The advent of deep learning has transformed recommendation systems. Neural collaborative filtering [3] combines matrix factorization with neural networks to learn non-linear user-item interactions. Knowledge graph embeddings [4] incorporate structured relationships between entities to improve recommendations.

### C. Pre-trained Language Models

BERT [5] revolutionized natural language processing by learning contextual word representations through masked language modeling. Sentence-BERT [6] adapted BERT for sentence-level embeddings using siamese networks. However, research has shown that vanilla BERT embeddings occupy a narrow cone in the embedding space (anisotropy), resulting in artificially high cosine similarities between unrelated texts [7].

### D. Contrastive Learning

Contrastive learning has emerged as a powerful self-supervised technique for learning discriminative representations. SimCLR [8] demonstrated that data augmentation combined with contrastive loss produces high-quality visual representations. Supervised contrastive learning [9] extends this framework by leveraging label information to define positive pairs.

## III. METHODOLOGY

In this paper, we build a course recommendation system by mapping student interest statements and course information into a shared semantic space and generating recommendations using similarity-based course ranking. The model builds on a pretrained BERT encoder (referred to as "Vanilla BERT" in this paper) and extends it with a self-supervised contrastive learning framework. Data augmentation is used during training to improve the model's consistency and its ability to capture semantic meaning, rather than relying on keyword matching or rule-based filtering. After projecting the tokenized embeddings into a lower-dimensional space, final course recommendations are selected using cosine similarity between the student input and course embeddings.

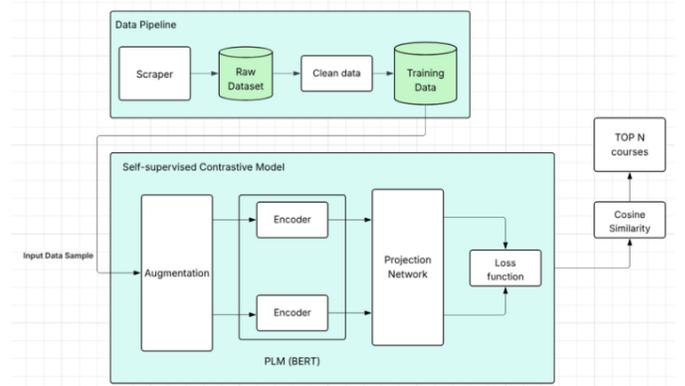

Fig. 2. Overall system flowchart

### E. Data Collection and Preprocessing

The data used in our application is divided into two main components: course data and synthetic student data.

#### 1) Course Data Collection:

Only engineering courses were used in our application, and course information was collected from the University of Ottawa course catalogue. For each course, the course code, title, description, components, and prerequisites were collected, but only the text-based fields were combined to produce a single semantic representation of each course.

TABLE I
SAMPLE CLEANED COURSE DATA

| Faculty | Code | TextforBERT |
|---------|------|-------------|
| ELG | 2136 | "electronics i. physics of semiconductors..." |

| MCG | 3110 | "heat transfer. fundamental principles of..." |
| CHG | 8301 | "renewable fuels. the production and sustainability of..." |

Text partially shown to account for space

Courses with sufficient descriptive content were kept in the dataset to ensure that only meaningful semantics were used. Since the University of Ottawa offers courses in English and French, only English courses were kept. This resulted in a total of 512 relevant engineering courses.

*2) Synthetic Student Statement Generation*

Student inputs were generated as common statements that could be used to express a student's liking of a specific subject. The structure of these inputs was made to vary the length and style of wording that might be used by students when naturally describing their interests. Therefore, the synthetic student data would look something like: "I enjoy learning about the processing of electrical power."

Each student input was associated with a single course considered to be a positive match. This tag would be used later during data augmentation and when evaluating the model. A total of 600 student statements were generated, and an 80/20 training/testing split was done.

TABLE II
SAMPLE SYNTHETIC STUDENT DATA

| Student Text | Liked Courses |
|---|---|
| i enjoy computer system design | CEG 3156; ELG 5383 |
| i like mechatronics | MCG 4136; ELG 6393 |
| i am interested in courses like fundamentals of geomechanics | CVG 7116 |

*3) Text Cleaning and Normalization*

The final part of the preprocessing was to ensure that all the text was normalized and clean. This step simply included lower casing and the removal of noninformative characters, as well as the normalization of white space. Overall, minimal filtering was applied to preserve the semantic structure. Both the course data and the synthetic student statements were processed through this cleaning pipeline to ensure effective tokenization by the pretrained BERT encoder.

*F. Augmentation*

Since our approach is based on a self-supervised contrastive learning framework, data augmentation is a required component of the training process. The primary motivation for this step is to provide the model with reliable positive pairs for each course, and to expand our effective data set, given that only 512 courses and 480 training student statements were available. Without data augmentation, the model wouldn't be able to learn, and semantic understanding would be limited.

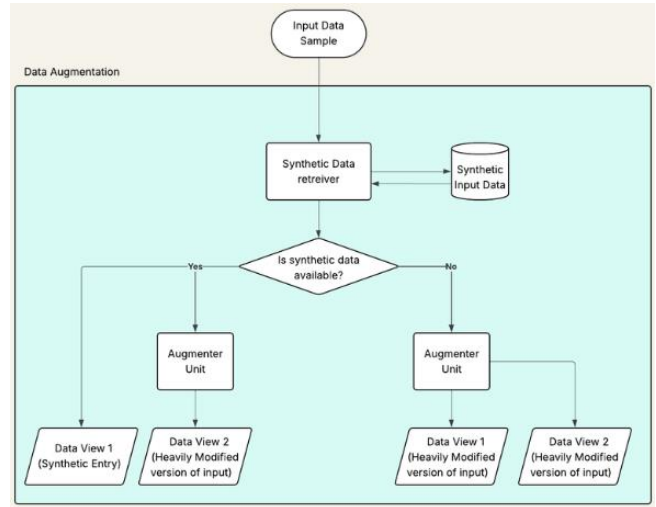

Fig. 3. Data augmentation process chart

The augmentation process follows the pipeline shown in Fig. 3. For each course data sample, a synthetic data retriever first checks whether an associated student statement exists. If available, the associated student statement is used as one view, while the second view is a heavily augmented version of the course data. If no associated student statement exists, both views are created through independent augmentations of the same course description. This design ensures that each training sample produces a valid positive pair corresponding to the same underlying course.

The following text-based augmentation methods were applied to course descriptions. Each augmentation operation was applied with different probabilities depending on whether a light or heavy augmentation was selected.

- Word Deletion
- Synonym Replacement
- Word Insertion
- Word Swapping

Text augmentation was implemented using the NLTK library, with WordNet used for synonym-based transformations. Augmentation was applied exclusively to course texts and only during training. Student statements were never modified and were used only when available to form positive pairs. This approach enables robust learning of course representations while incorporating real student intent when available, improving generalization in similarity-based recommendations.

*G. Model Base*

Our model is built on a pretrained BERT backbone and is extended with a small projection head to produce embeddings suitable for similarity-based course recommendations. BERT processes input text by first splitting it into tokens (text units produced by the tokenizer), where each token is mapped to a numerical vector whose representation depends on the surrounding context. This allows the model to capture semantic meaning, not just individual words. The encoder and projection head were implemented using PyTorch, with pretrained BERT weights loaded through the Hugging Face Transformers library.

Since the recommendation pipeline requires a single vector per input, masked mean pooling is applied over the final hidden states, which are the vector output of the final BERT layer. This operation averages the token vectors across the sequence while ignoring padding tokens. The attention mask (a binary indicator for valid tokens) ensures that only meaningful tokens contribute to the pooled representation. The result is a fixed size embedding that represents the overall semantic content of the input and remains robust to variations in input length.

To adapt the encoder for contrastive training, the pooled embedding is passed through a projection head composed of two fully connected layers with a ReLU activation in between. The first layer operates on the original 768-dimensional pooled BERT representation and introduces nonlinearity, allowing the model to learn weights for the semantic features extracted by BERT. The second layer projects this representation onto a lower-dimensional embedding space (256 dimensions in our implementation) and is used for similarity-based comparison. The resulting embedding $z \in \mathbb{R}^{256}$ is L2 normalized so that similarity comparisons depend only on vector direction rather than magnitude. L2 normalization ensures that both the contrastive loss during training and cosine similarity during inference depend only on embedding direction rather than magnitude.

### H. Loss Function

Training is guided by a contrastive loss that enforces semantic similarity in the learned embedding space. The loss operates on embeddings produced from the two views generated by the encoder. Although the overall framework follows a self-supervised contrastive learning approach based on view generation and data augmentation, the training uses a supervised contrastive loss where positives are defined using available labels.

For a batch of size $B$, the encoder produces two L2-normalized embeddings per sample, corresponding to the two views used in contrastive learning. These embeddings are concatenated to form a single matrix, $Z \in \mathbb{R}^{2B \times D}$, which serves as the input to the contrastive loss. Similarity is computed between all pairs of embeddings within the batch, allowing the loss to use the predefined positive samples while treating the remaining embeddings in the batch as negatives. While based on the standard NT-Xent formula, the loss is extended to a supervised, multi-positive setting in which all samples sharing the same label are treated as positives. The NT-Xent loss function is defined as:

$$\mathcal{L}_a = -\log\left(\frac{\sum_{p \in P(a)} \exp\left(\frac{\text{sim}(Z_a, Z_p)}{\tau}\right)}{\sum_{k \neq a} \exp\left(\frac{\text{sim}(Z_a, Z_k)}{\tau}\right)}\right) \quad [1]$$

where:
- $Z \in \mathbb{R}^{2B \times D}$ is the matrix of all embeddings in the batch
- $Z_a$ is the current embedding (one row of Z)
- $P(a)$ is the set of positive samples for the current embedding, defined as all embeddings sharing the same label and excluding the embedding itself.
- sim(A, B) denotes cosine similarity where:
$$\text{sim}(A, B) \coloneqq \cos(\theta) = \frac{\mathbf{A} \cdot \mathbf{B}}{|\mathbf{A}||\mathbf{B}|} \quad [2]$$
- $\tau$ is a temperature parameter controlling the sharpness of the similarity distribution.
- B is the batch size, and D is the embedding dimension.

The loss is computed for all the embeddings in a batch and averaged across that batch. The temperature parameter was tuned empirically to provide improved isotropy without sacrificing accuracy of the model.

In addition to the contrastive loss, an isotropy loss term is applied to the embeddings. This loss encourages zero-mean and unit-variance features across the batch, meaning a more uniform distribution in the embedding space. The final training loss is given by:
$$\mathcal{L} = \mathcal{L}_{\text{cont}} + \lambda \mathcal{L}_{\text{iso}}$$
Where $\lambda$ controls the contribution of the isotropy loss.

The model was trained using the AdamW optimizer with a linear learning rate schedule and warm-up. Training was performed for a fixed number of epochs using mini-batches, and gradient clipping was applied to improve stability. All hyperparameters were selected empirically and kept fixed across experiments.

## IV. RESULTS

We begin by illustrating the recommendation process using a sample student input. Given a student interest statement, the trained model maps the input into the same embedding space as the course representations. The recommendations are then generated by ranking courses using cosine similarity. For the example shown, the top-N (top-5 in our implementation) recommended courses align well with the student's stated interests, despite differences in wording between the input and course descriptions.

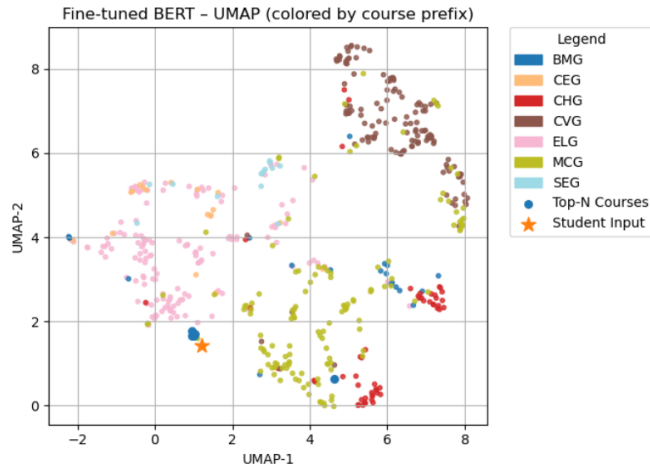

Fig. 4. Sample output with input example

A two–dimensional UMAP (Uniform Manifold Approximation and Projection) illustrates the model's behavior, where the student's embedding is close to its top-ranked courses. This provides an intuitive visualization of how recommendations are produced in the learned semantic space.

Fig. 5. UMAP output showcasing all courses mapped for the fine-tuned BERT model

### I. Recommendation Accuracy Evaluation

Evaluation of the recommender system was performed on a test student input dataset, where statements were embedded and matched against the full set of course embeddings. Performance was measured using Hit Rate, F1 score, and Mean Reciprocal Rank (MRR), which evaluate the top N retrieval accuracy, ranking quality, and early relevance, respectively. In addition to these common recommender metrics, embedding isotropy was analyzed separately using the IsoScore [10] metric to assess the geometric quality of the learned representation space.

Table III reports the performance of Vanilla BERT and the proposed contrastive model. Vanilla BERT exhibits low retrieval performance across all metrics, indicating that the pretrained embeddings are weak for a semantic course recommendation application. In contrast, our model achieves substantially higher scores across all metrics, highlighting the effectiveness of our contrastive, self-supervised approach. Also, among the evaluated temperature values, a temperature of $\tau = 0.05$ yields the best overall performance, indicating an effective balance between positive alignment and negative separation.

TABLE III
METRICS

| Metric | Vanilla BERT | Our Model | | |
|---|---|---|---|---|
| | | $\tau = 0.2$ | $\tau = 0.05$ | $\tau = 0.01$ |
| Hit Rate | 0.033 | 0.833 | 0.917 | 0.925 |
| F1 Score | 0.021 | 0.654 | 0.725 | 0.733 |
| MRR | 0.014 | 0.658 | 0.733 | 0.76 |
| IsoScore | 0.818 | 0.031 | 0.065 | 0.082 |

### J. Embedding Structure and Isotropy

The performance gains observed in section A are further supported by analyzing the geometric structure of the learned embedding space. Fig. 6. compares UMAP projections of course embeddings produced by Vanilla BERT and the contrastively trained model. Vanilla BERT embeddings exhibit strong anisotropy, with many courses clustered closely together regardless of semantic differences. In contrast, the trained model produces a more evenly distributed embedding space with clearer separation between course groups, enabling more meaningful similarity comparisons.

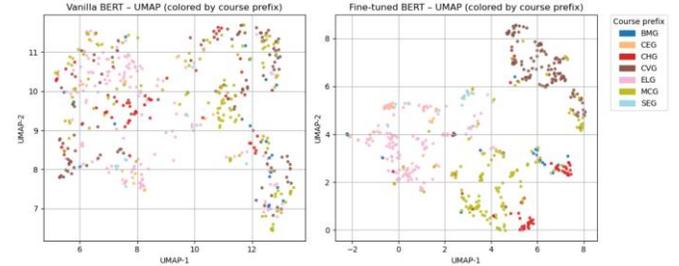

Fig. 6. Comparison of UMAP output for vanilla BERT and fine-tuned BERT

Table IV reports the mean and standard deviation of cosine similarity values between randomly sampled course pairs. Vanilla BERT embeddings show a high mean similarity with low variance, indicating limited discriminability. After contrastive training, the mean similarity decreases and variance increases, reflecting improved separation between unrelated courses. This trend is further confirmed by the IsoScore metric, which reports a higher isotropy score for the trained model.

TABLE IV
COURSE COSINE SIMILARITY DISTRIBUTION

| Cosine similarity | Vanilla BERT | Our Model | | |
|---|---|---|---|---|
| | | $\tau = 0.2$ | $\tau = 0.05$ | $\tau = 0.01$ |
| Mean | 0.818 | 0.016 | 0.115 | 0.469 |

| Standard deviation | 0.054 | 0.271 | 0.19 | 0.135 |

The effect of the temperature parameter on embedding structure is illustrated in Fig. 7. Higher temperature values reduce embedding separation, while excessively low temperatures produce overly compact or unstable representations. A temperature of τ = 0.05 produces the most balanced structure, which aligns with the highest recommendation performance reported in Table III.

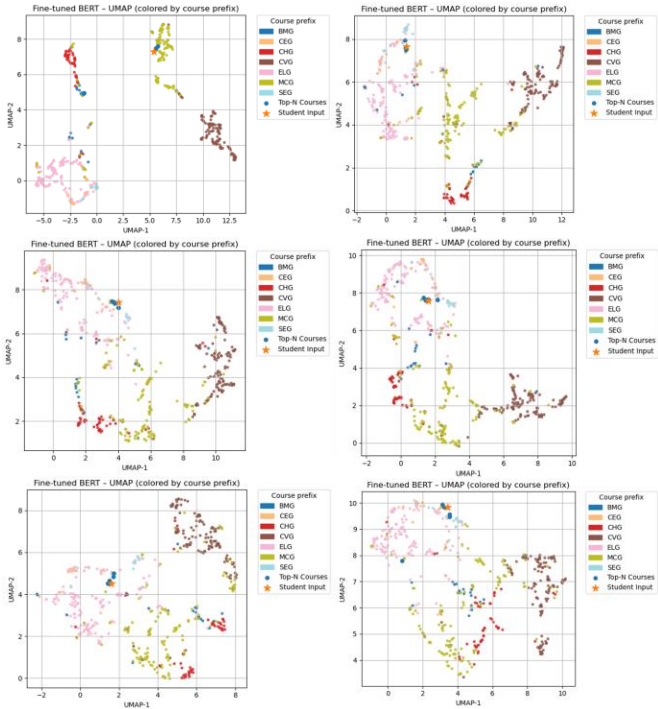

Fig. 7. Collection of all UMAP outputs for models trained with varying parameters

## V. CONCLUSION

In this work, we presented a course recommendation system for engineering students based on a self-supervised contrastive learning framework built on top of a pretrained BERT encoder. The proposed approach maps free-form student interest statements and course descriptions into a shared semantic embedding space and generates recommendations using cosine similarity–based ranking. By focusing on semantic understanding rather than historical enrollment data, the system effectively addresses common challenges in academic recommendation, including cold-start scenarios and niche student interests.

A key limitation of vanilla BERT embeddings—anisotropy in the representation space—was identified as a major obstacle for similarity-based course recommendation. To mitigate this issue, we introduced contrastive learning with text-based data augmentation and incorporated an isotropy regularization term to improve the geometric structure of the embedding space. Experimental results demonstrate that the proposed model produces more discriminative and evenly distributed embeddings, leading to substantial improvements in recommendation accuracy compared to vanilla BERT baselines.

Quantitative evaluation using Hit Rate, F1 score, and Mean Reciprocal Rank confirms that the contrastively trained model significantly outperforms pretrained BERT embeddings across all retrieval metrics. Additional analysis using IsoScore and cosine similarity distributions further validates that the learned embedding space exhibits improved isotropy and separation between unrelated courses. Visualization using UMAP projections provides intuitive evidence of these improvements, showing clearer clustering and alignment between student inputs and relevant courses.

While the proposed system demonstrates strong performance, several directions for future work remain. Incorporating real student interaction data could enable personalized recommendations over time, while extending the dataset beyond engineering courses would allow cross-faculty recommendations. Additionally, exploring larger language models or interactive feedback mechanisms may further improve recommendation quality and usability. Overall, this work highlights the effectiveness of contrastive learning for semantic course recommendations and demonstrates its potential as a practical decision-support tool in academic advising systems.